\documentstyle[psfig]{aa}     

\newcommand{\Lsun} {L$_\odot$}
\newcommand{\Msun} {M$_\odot$}
\newcommand{\um} {$\mu$m}

\begin{document}
\thesaurus{06 (09.04.1, 08.16.5, 08.09.2, 03.20.5, 13.09.3) }

\title{The protostellar system HH108MMS\,\thanks{Based on observations
with ISO, an ESA project with instruments funded by ESA Member States
(especially the PI countries: France, Germany, the Netherlands and the
United Kingdom) with the participation of ISAS and NASA.}}

\author {R.~Siebenmorgen\inst{1} and E.Kr\"ugel\inst{2}}

\institute{ 
	\inst{1}  European Southern Observatory,
	 Karl-Schwarzschildstr. 2,
	 85748 Garching b. M\"unchen, Germany \\
	\inst{2} MPIfR, Auf dem H\"ugel 69, D-53121 Bonn }

\offprints{rsiebenm@eso.org}
\date{Received 12 May 2000; accepted 20 October 2000}
\authorrunning {Siebenmorgen \& Kr\"ugel}
\titlerunning {The protostellar system HH108MMS}

\maketitle

\begin{abstract}
We probe the region around the protostar HH108MMS by deep mid infrared
photometric and polarimetric imaging.  The protostar is detected at
14$\mu$m in absorption against the diffuse background.  Next to
HH108MMS, we find a second absorbing core, named Q1, and the young
stellar object IRAS18331--0035 which is more advanced in its evolution
and already seen in emission at 12$\mu$m and 14$\mu$m.  HH108MMS, Q1
and IRAS18331--0035 form a triplet along an extended filamentary
absorption feature.  From the variation of the surface brightness
across the source, we derive for HH108MMS and Q1 the optical depth and
density profile.  Along the axes which are parallel to the filament,
the density distributions follow a $\rho\propto r^{-1.8}$ power law.
We estimate that the intensity of the background radiation at 14$\mu$m
is about two times stronger than the intensity of the interstellar
radiation field in the solar neighborhood. The present photometric
data of IRAS18331--0035 between 12$\mu$m and 1.3mm can be explained by
a central source with a luminosity of 2.5\,\Lsun\ that is surrounded
by a spherical cloud of 1.1\,\Msun\ with a $1/r$ density
distribution. As HH108MMS is also seen in the millimeter dust
emission, we can derive the ratio of the dust extinction coefficients
at 14$\mu$m and 1.3mm and obtain $\kappa_{14\mu\rm m} /
\kappa_{1300\mu \rm m}\sim 470$.  Because models for the dust in the
diffuse interstellar medium predict a ratio of around 2000, our value
points to fluffy composite grains which are expected to prevail in
dense and cold environments. 

\noindent First mid infrared polarisation images of pre--stellar
absorbing cores are presented.  At 12$\mu \rm{m}$ and 14$\mu \rm{m}$
the polarisation of the region around HH108MMS is strong ($\geq 15\%$)
and tightly correlated with the source triplet.  We demonstrate that
the high degree of polarisation can be explained by extinction of
rotationally aligned dust particles of moderate elongation.

\keywords{ISM: dust, extinction -- Stars: pre-main sequence -- Stars:
individual: HH108MMS -- Techniques: polarimetric -- Infrared: ISM:
continuum}
\end{abstract}

\section{Introduction}
\noindent 
Herbig Haro objects mark shock regions in the outflow from very young
stars.  The 1.3mm continuum survey of Herbig Haro objects by Reipurth
et al.~(1993) indicates that the young stars, which are the energy
sources of the outflow, are surrounded by dusty envelopes of about 0.1
-- 3 solar masses.  The circumstellar material is so massive that the
stars are probably still in the accretion phase.  In a follow--up
study of the 1.3mm dust continuum, Chini et al. (1997) discovered in
the Serpens star forming region which is at a distance $D= 310$\,pc
(De Lara et al. 1991) a protostellar candidate located 71$''$
(0.11pc) away from the source IRAS18331--0035.  The latter is believed
to be the driving engine of the two aligned nearby bow shocks HH108
and HH109 (Reipurth \& Eiora 1992, Ziener \& Eisl\"offel 1999); they
lie to the South-West of IRAS18331--0035, about 0.1pc and 0.21pc away.
In this paper, we present photometric and polarimetric mid infrared
images of the region.

\section{Observations}

\noindent 
{\it 2.1 Photometric imaging} \\

\begin{figure*}
 
\centerline{ \hspace{4.0cm} {\psfig{figure=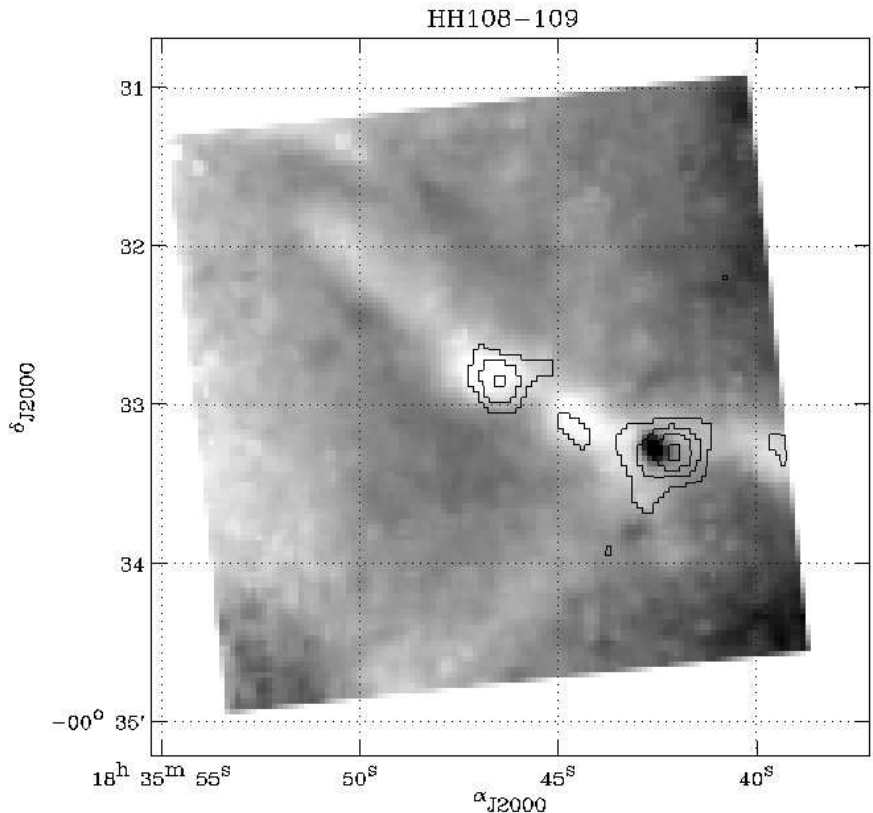}} }
\caption{ ISOCAM lw3 image (14.3$\mu$m) in grey scale around the
protostar HH108MMS (30$'' \sim 9300$AU).  The contour lines are an
overlay from Chini et al.~(1997) and show the 1.3mm dust emission.  In
the mid infrared, the right 1.3mm source, IRAS18331--0035, is in
emission, the left, HH108MMS, is seen in absorption.  The other
absorbing knot in between is denoted Q1; it is weak at 1.3mm, but
clearly detected. There is a hint of a third absorbing core in the
uppermost West.  The ISOCAM image shows a large filamentary structure
into which HH108MMS, Q1 and IRAS18331--0035 are embedded.  }
\label{HH108_tempflat_bw_230799.ps}
\end{figure*} 

\noindent 
The protostellar system HH108MMS was photometrically imaged on 5 March
1996 with the long wavelength array of ISOCAM (Cesarsky et al.~1996)
on board the infrared space observatory (ISO).  The morphology of the
object was studied by applying the observing template CAM01 in raster
mode.  A 5 $\times$ 5 micro-raster was performed with step size of
20$\arcsec$ in both space craft directions; we used the broad band
filter lw3 (12 -- 18 \um), the 6$\arcsec $ lens and the large field
mirror.  After 100 stabilization frames, 33 exposures were read out at
each raster position.  The read--out time for each exposure was 2.1s.

\noindent 
The data were reduced with the ISOCAM interactive analysis (CIA). 
The basic reduction
steps such as dark current subtraction, initial removal of cosmic ray
hits, detector transient fitting and photometry are described in
Siebenmorgen et al.~(1999).  Because in our observing strategy several
detector elements saw the same part of the sky at different times, we
used the redundant information to correct for the long term camera
drift and to fine--tune the deglitching process (Miville--Deschenes et
al.  2000).  The coadded images at each raster position were then
sky--projected and corrected for field distortion.  The final raster
map has a pixel scale of 2$\arcsec$ which corresponds to 
$\sim$\,620AU at the distance of Serpens ($D= 310$\,pc).

\noindent 
Fig.~\ref{HH108_tempflat_bw_230799.ps} displays in grey scale our lw3
image of HH108MMS and its surroundings.  Superimposed are contours of
the 1.3mm emission as observed by Chini et al.~(1997).  The brighter
1.3mm source coincides with IRAS\,18331--0035.  At 14$\mu\rm m$ we
detect this source in emission, HH108MMS itself is seen in absorption.
Besides a second absorption core, Q1, and a tentative third absorbing
knot, we find a filamentary structure again in absorption against the
diffuse infrared background.  The filament extends about 200\arcsec \/
(0.3 pc) along its major and 20\arcsec \/ (0.03 pc) along its minor
axis.  The major axis is oriented at a position angle PA = 45$^o$.

\noindent 
Photometry of IRAS18331--0035 in a 12$''$ aperture gives $F_{14\mu\rm
m} = 4.2 \pm 0.3$\,mJy. The fluxes have been scaled by a correction
factor for a monochromatic point spread function and color--corrected
assuming a $F_\lambda \propto\lambda$ spectrum in the band. \\

\begin{figure}
\centerline{  \hspace{2.5cm}
\psfig{figure=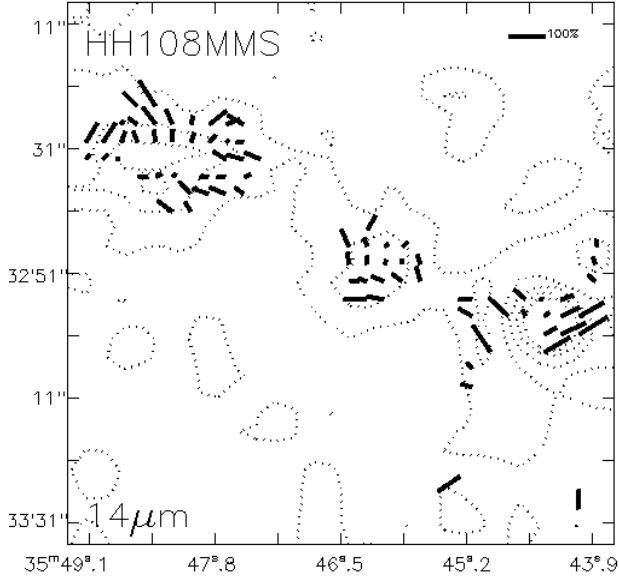,width=11.50cm} }
\caption{Polarisation map at 14.3$\mu$m (lw3) of the protostellar
system HH108MMS (cmp. with Fig.~1). The intensities are shown as
contour overlay and polarisation vectors as bars.  For the absorbing
cores the polarisation vectors indicate the magnetic field direction.}
\label{hh108_polmapp_lw3_bw.ps}
\end{figure}

\begin{figure}
\centerline{  \hspace{2.5cm}
\psfig{figure=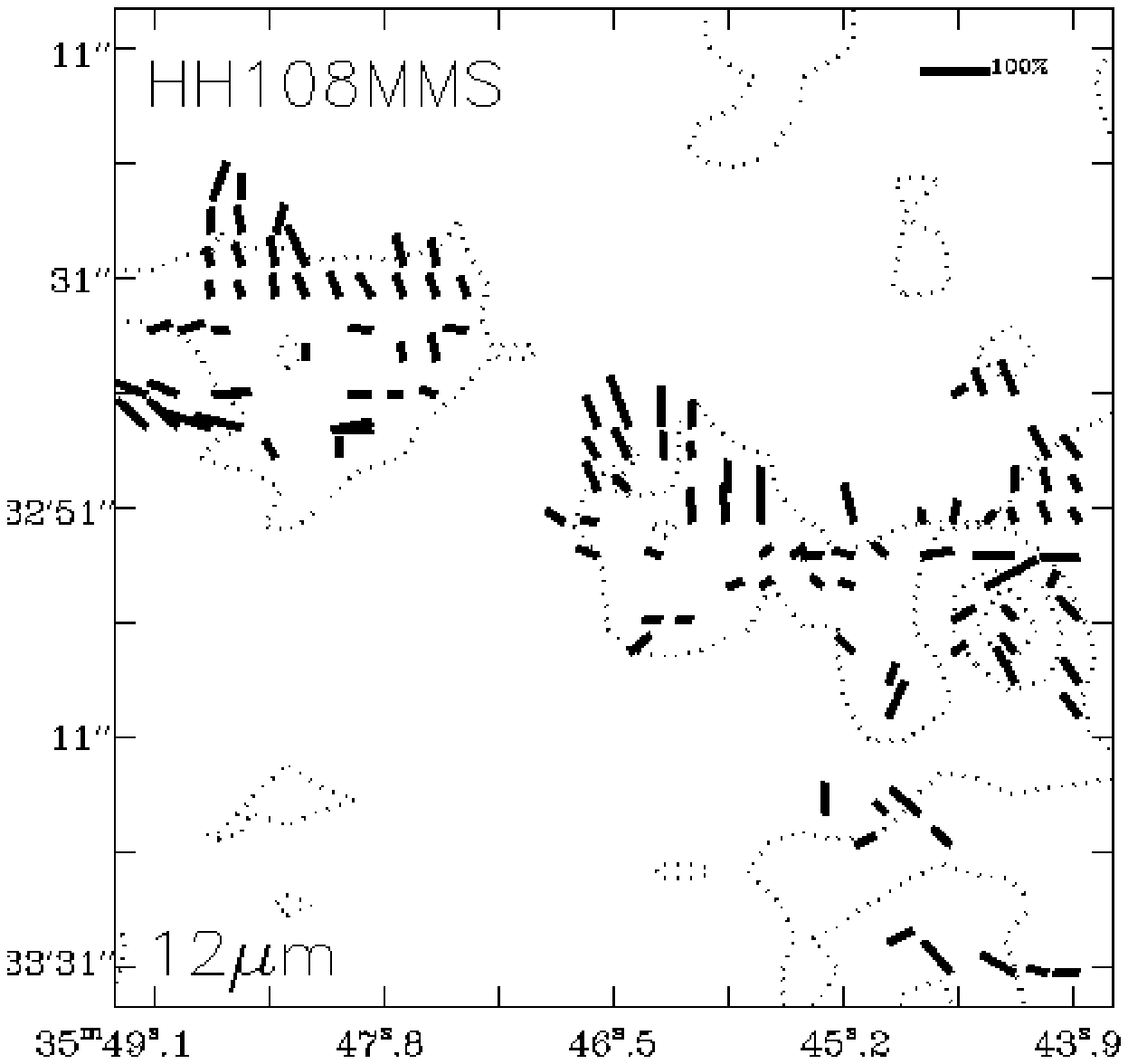,width=11.50cm}} 
\caption{As Fig.~\ref{hh108_polmapp_lw3_bw.ps} for filter lw10 (12.0 $\mu$m).}
\label{hh108_polmapp_lw10_bw.ps}
\end{figure}

\begin{figure}
\psfig{figure=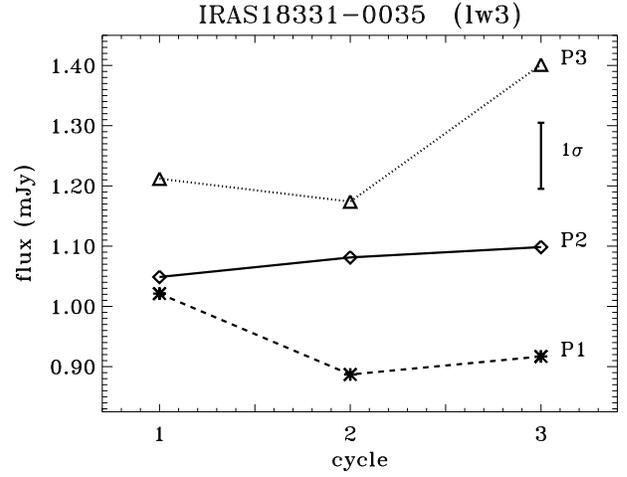,width=9.0cm,angle=90} 
\caption{Flux of IRAS18331-0035 as measured through filter lw3 (14.3
$\mu$m) and polariser P1 (*), polariser P2 ($\diamondsuit$) and
polariser P3 ($\bigtriangleup$) for the different observing
cycles. The typical 1$\sigma$ error bar is indicated. } 
\label{IRAS18331_pol_lw3.ps}
\end{figure}

\begin{figure}
\psfig{figure=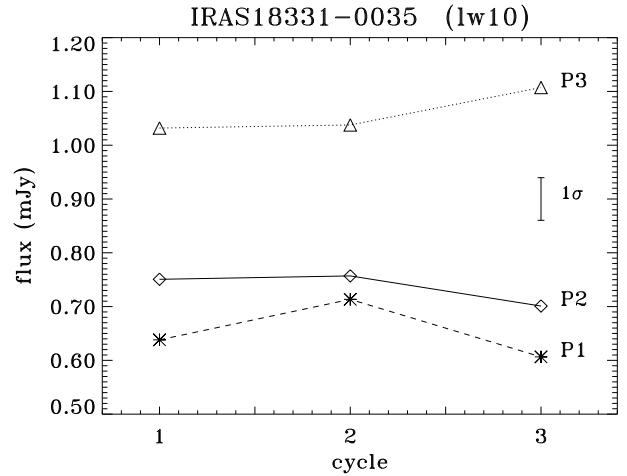,width=9.0cm,angle=90} 
\caption{As Fig.~\ref{IRAS18331_pol_lw3.ps} for filter lw10 (12.0
$\mu$m).}
\label{IRAS18331_pol_lw10.ps}
\end{figure}

\noindent 
{\it 2.2 Polarimetric imaging} \\

\noindent 
Polarimetric images of the HH108MMS region were obtained on 30 March
1998 with ISOCAM.  The polarisation pattern was studied by applying
the observing template CAM05 (Siebenmorgen 1996) in broad band filters
lw3 (14.3 $\mu$m) and lw10 (12.0 $\mu$m).  In each filter, a 2
$\times$ 2 micro-raster was performed at a step size of 54$\arcsec$ in
both space craft directions.  The camera was configured in the
6$\arcsec $ lens and the large field mirror.  After 30 stabilization
frames, 15 exposures were read out at each raster position.  The
read--out time for each exposure was 5s at gain two.  After a raster
without polarisers, similar rasters were taken for each of the three
polarisers.  The polariser rasters were repeated over three observing
cycles.  The nominal pointing position (18$^{\rm h}$ 35$^{\rm m}$
46$^{\rm s}$.5,-- 0$^{\circ}$ 32$'$ 51$''$, J2000.0) of the target was
corrected in the polariser rasters for the known source displacement
on the detector.

\noindent 
The polarisation maps of HH108MMS in filters lw3 and lw10 are shown in
Fig.~\ref{hh108_polmapp_lw3_bw.ps} and
Fig.~\ref{hh108_polmapp_lw10_bw.ps}.  Superimposed are contours of the
intensities without the polarisers.  They reveal the same structure as
detected in the CAM01 observations.  Photometry of IRAS18331-0035 at
12 $\mu$m (lw10) gives $2.0\pm 0.2$\,mJy and at 14 $\mu$m (lw3) $4.2
\pm 0.4$\,mJy, in accordance with the CAM01 observations.

\noindent 
As in the CAM01 observations described above, we applied the basic
data reduction steps, such as dark current subtraction, initial
removal of cosmic ray hits and transient correction.  A flat field is
derived by taking advantage of the fact that all detector pixels spend
more time observing the background than on the source.  The coadded
images at each raster position were projected on the sky and corrected
for field distortion to derive the final mosaics.  The mosaics have a
pixel scale of 3$''$.  The different raster maps of the same polariser
are registered on the brightest pixel of IRAS18331--0035.  We measure
a relative shift of the source of $\leq 1$~pixel from one cycle to the
next.  The polarised signal is found to be consistent between the
cycles, so that an average of all cycles gives the final mosaic image
for each polarisers.

\noindent 
From each mosaic we subtract the background.  Its value is estimated
on the same sky area. The Stokes elements are calculated following the
notation and calibration factors given by Siebenmorgen (1999).  The
background at 12$\mu$m and 14$\mu$m is mostly due to zodiacal light.
By assuming the latter to be unpolarised, we get an internal
calibration of the instrumental polarisation which is consistent with
the numbers cited in Siebenmorgen (1999).

\noindent 
Fig.~\ref{IRAS18331_pol_lw3.ps} displays the 14$\mu$m fluxes of
IRAS18331--0035 as measured through the three polarisers P1, P2 and P3
over three observing cycles.  For a given polariser, the fluxes agree
within the uncertainty limits.  However, the different polarisers have
different fluxes, so the mid infrared emission of IRAS18331--0035 is
polarised.  Fig.~\ref{IRAS18331_pol_lw10.ps} is similar and refers to
the 12$\mu$m filter.  Altogether we find:
\begin{itemize}
\item [--] 12$\mu$m: \  $p = 28.9\pm 4.6 \%$, $\theta = 115 \pm 5^o$. 
\item [--] 14$\mu$m: \  $p = 15.2\pm 4.8 \%$, $\theta = 108 \pm 9^o$. 
\end{itemize}

\begin{figure} 
\centerline{  \vspace{-1.3450cm}  
\psfig{figure=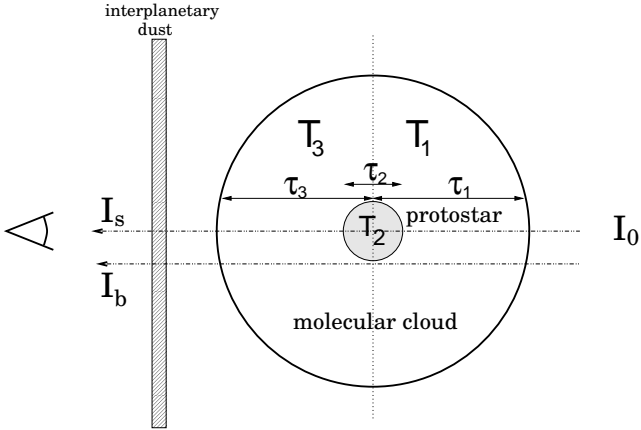,width=11cm,angle=-90}}
\caption{A sketch showing how we envisage that HH108\,MMS is embedded
into the molecular cloud.  We indicate the temperature and optical
depth in the various regions.  Looking towards HH108\,MMS, one
observes the intensity $I_{\rm s}$, away from it $I_{\rm b}$; \ $I_0$
and $I_{\rm z}$ are the intensity behind the Serperns cloud and of
the foreground zodiacal light, respectively.  }
 \label {intensity_region.ps}
\end{figure}

\noindent 
Repeating the observing procedure on either an unpolarised standard
star (HIC085371) or the zodiacal light, the fluxes through all
polarisers are implying that the instrumental polarisation is smaller
than 1.5\% (Siebenmorgen 1999).

\section{Results}

\noindent 
{\it 3.1 The optical depth profiles of the protostars} \\

\begin{figure} 
\centerline{ \hspace{-0.75cm}\psfig{figure=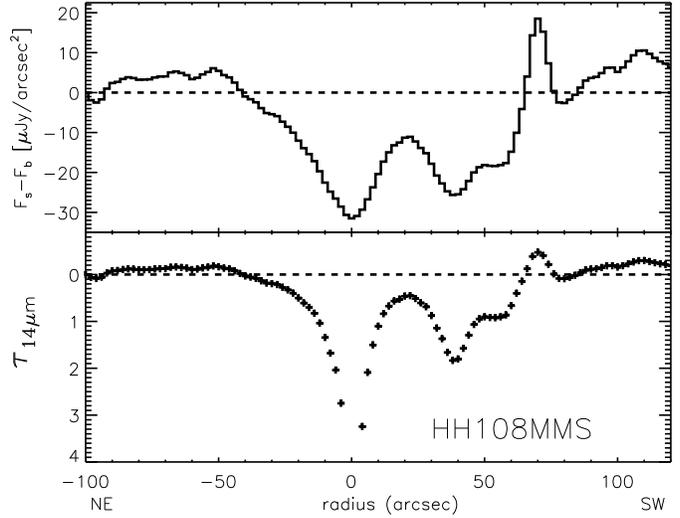,width=9.25cm,bbllx=65bp,bblly=695bp,bburx=535bp,bbury=85bp,clip=,angle=90}}
\caption{ {\it Top:} The difference between the flux towards
HH108\,MMS and the adjacent background in a cut along the major axis.
The major axis goes through the position of maximum absorption in
HH108\,MMS, where the offset is zero and $\tau_2 = \tau_{\rm 2,max}$,
along a line connecting to the center of Q1.  {\it Bottom:} The
optical depth is derived after Eq.(\protect\ref{Gl5}).  }
\label{tau_major.ps}
\end{figure}

\noindent 
To derive the density structure of the protostellar systems,
HH108\,MMS and Q1, we transform the mid infrared surface brightness
into an optical depth.  The underlying geometrical configuration is
depicted in Fig.~\ref{intensity_region.ps}.  HH108\,MMS is embedded in
the Serpens molecular cloud and viewed from Earth through a veil of
zodiacal light.  We distinguish four regions that contribute to the
observed flux:

\begin{itemize}
\item 
The background with radiation of intensity $I_0$ that heats the
molecular cloud from behind.
\item 
The far and near side of the molecular cloud.  They have constant
optical depths $\tau_1, \tau_3$ and constant temperatures $T_1, T_3$,
respectively.
\item 
The absorbing core itself with corresponding parameters $\tau_2$ and
$T_2$.
\item 
The interplanetary dust which is tenuous and emits zodiacal light of
intensity $I_{\rm z}$.
\end{itemize}

\noindent
To simplify the analysis, we assume:

\begin{itemize}
\item 
$\tau_1 = \tau_3$, the absorbing core is located in the optical center
of the molecular cloud;
\item 
$T_1 = T_3$, the temperature is uniform over the molecular cloud;
\item 
$B_{14\mu \rm{m}}(T_1) = 0$, the molecular cloud is cold and has no PAH emission;
\item 
$B_{14\mu \rm{m}}(T_2) = 0$, the absorbing core is isothermal and has also no mid infrared
emission of its own.
\end{itemize}

\noindent
In the direction of the absorbing core (source) we can thus write

\begin{equation}\label{Gl1}
I_{\rm s} \ = \ I_{\rm z} \ + \ I_0 \cdot e^{-(2\tau_1+\tau_2) }
\end{equation}

\noindent 
Looking at a position next to the absorbing core, one receives the
background intensity

\begin{equation}\label{Gl2}
I_{\rm b} \ = \ I_{\rm z} \ + \ I_0 \, e^{-2\tau_1}
\end{equation}

\noindent
If the zodiacal light is constant on scales of a few arcsec, this leads to

\begin{equation}\label{Gl3}
I_{\rm s} - I_{\rm b}  \ =  \ I_0 \cdot e^{-2\tau_1} \cdot [e^{-\tau_2} - 1] \ .
\end{equation}

\noindent
Of course, the optical depth $\tau_2$ varies across the absorbing
core.  Let us denote its maximum value by $\tau_{\rm 2,max}$ and by
$I_{\rm s,max}$ the observed intensity towards that position.  An
estimate for $\tau_{\rm 2,max}$ is obtained from the 1.3mm peak flux
in a 11$''$ beam of $S_{1.3\rm mm} = 175$\,mJy (Chini et al.~1997) and
the dust temperature of 13\,K (Chini et al.~2000). From the formula
$S_\nu = B_\nu(T) \,\Omega\, \tau_\nu$ with a solid angle $\Omega =
2.2\cdot 10^{-9}$, we find that $\tau_{1.3\rm mm} \sim 7\cdot
10^{-3}$.  This is an average over the 11$''$ beam.  As the source is
centrally peaked at 1.3mm, we expect the optical depth in a $2''\times
2''$ area, corresponding to the pixel size of the ISOCAM map, to be a
few times larger.  When we adopt a dust extinction coefficient of
$\kappa_{1.3\rm mm}= 0.02\, {\rm cm}^2$ per g of interstellar matter,
as would be appropriate for a class~0 object (Kr\"ugel \& Siebenmorgen
1994), and the total 1.3\,mm flux of 282\,mJy, we derive a total gas
mass of 0.5\Msun.

\noindent 
To convert $\tau_{1.3\rm mm}$ into the maximum optical depth at 14$\mu$m,
$\tau_{\rm 2,max}$, we employ the dust models by Kr\"ugel \& Siebenmorgen
(1994).  For the diffuse interstellar medium, we read off from their Fig.~12 an
optical depth ratio $\tau_{14\mu \rm m}/\tau_{1.3\rm mm}\sim 1900$.  For grains
of the same size in cold and dense cores where they have ice mantles and a
fluffy structure, the ratio is about 1000.  We thus end up with an estimate for
$\tau_{\rm 2,max}$ that is above 10 and thus comfortably above unity.  We may
now safely assume that $\exp\,(-\tau_{\rm 2,max})$ is small so that (see Eq.(3)

\begin{equation}\label{Gl4}
I_0 \cdot e^{-2\tau_1} = I_{\rm b} - I_{\rm s,max}  \ .
\end{equation}

\noindent
Inserting $I_0e^{-2\tau_1}$ into Eq.(\ref{Gl3}) gives the following
approximate expression for the variation of the optical depth in a cut
through the major axis of the core,

\begin{equation}\label{Gl5}
\tau_2 = - \ln\left(1- {I_{\rm b}-I_{\rm s}\over I_{\rm b} - I_{\rm s,max}}
\right) \ .
\end{equation}

\noindent 
It is independent of the optical depth of the molecular cloud,
$2\tau_1$.  The background intensity $I_{\rm b}$ is determined as an
average over two strips that run parallel to the major axis, one
above, the other below it; we find $I_{\rm b} =
4.3 \cdot 10^{-16}$\,erg/s cm$^2$ Hz ster.

\noindent 
We can also derive $I_0$ from Eq.(\ref{Gl4}) and thus the rate at
which the cloud is heated from outside if we can guess $\tau_1$.  A
rough number for $2\tau_1$ in Eq.(\ref{Gl4}) comes from the H$_2$CO
line observations in the Serpens cloud by Ungerechts \& G\"usten
(1984) who propose a visual extinction of 16\,mag.  With a ratio of
$\tau_{\rm V}/ \tau_{14\mu\rm m} \sim 50$ for dust in the diffuse
medium, one obtains $2 \tau_1 \sim 0.3$.  Our ISOCAM data then yield
an intensity at 14$\mu$m: $I_0 = e^{0.3} \cdot
30\mu$\,Jy\,arcsec$^{-2}$ or \\

$I_0 (14\mu \rm{m}) = 1.7\cdot 10^{-17}$\,erg/s cm$^2$ Hz ster, \\

\noindent
which is about two times higher than in the solar vicinity where
$I_{\rm ISRF}(14\mu\rm {m}) = 7\cdot 10^{-18}$\,\,erg/s cm$^2$ Hz ster
(Perault 1987).  So the value of $I_0$ is reasonable considering that
it refers to a star forming region where the radiation field is
stronger.

\noindent 
The combination of the ISOCAM and 1.3mm images allows to estimate the
extinction ratio $\kappa_{14\mu\rm{m}} / \kappa_{1.3\rm{mm}}$.  When
we roughly evaluate from the ISOCAM image the average optical depth of
the absorbing core over the 11$^{''}$ beam of the 1.3mm map, we find
for this average $\overline{\tau_2} \sim 3$ (Fig.~\ref{tau_major.ps}).
Therefore,
\begin{equation}\label{Gl6}
{ \kappa_{14\mu\rm m}\over \kappa_{1.3\rm mm}} \ = \ {(\overline{\tau_2} +
2\tau_1)_{14\mu\rm m} \over\tau_{1.3\rm mm}} \ \sim \ 470 
\end{equation}

\vskip0.5cm
\noindent 
{\it 3.2 The density structure of the protostars} \\

\noindent 
From the variation of the optical depth over the source
(Fig.~\ref{tau_major.ps}) one can derive the density structure of the
protostellar core.  Consider a spherical cloud of Radius $R$ and density
$\rho(r) = \rho_0\,r^{-\alpha}$.  The optical depth $\tau_2(x)$ through the
cloud at an offset $x\le R$ from its center is given by

\begin{equation}\label{Gl7}
\tau_2(x) = \kappa\,\rho_0 \int\limits^{\sqrt{R^2-x^2}}_0
(x^2+s^2)^{-\alpha/2}\,ds
\end{equation}

\noindent 
where $\kappa$ is the dust absorption coefficient.  For determining the exponent
$\alpha$ of the density distribution, the product $\kappa\rho_0$ is an
irrelevant proportionality factor.  The value of $\tau_2(x)$ depends on the
upper boundary of the integral and thus on $R$. From the upper panel of
Fig.~\ref{tau_major.ps}, we estimate a source size for HH108MMS of 60$''$ which
corresponds to a cloud radius $R=10000$\,AU.

\begin{figure} [htb]
\centerline{\psfig{figure=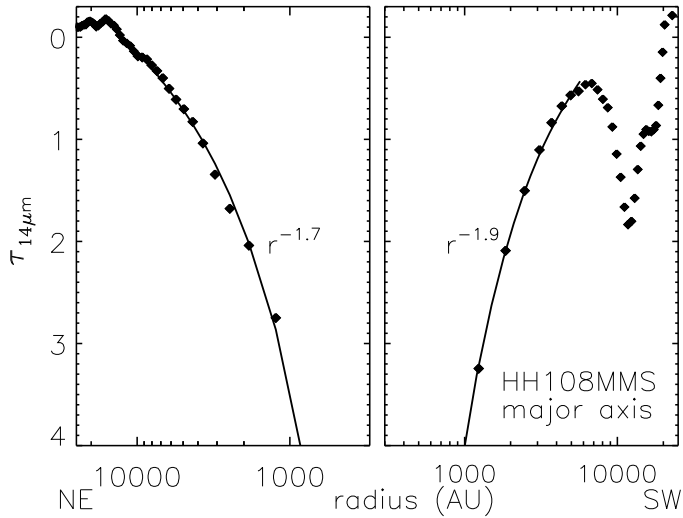,width=10cm,angle=90}}
\caption{ The optical depth profile along the major
axis of HH108MMS as from Fig.~\ref{tau_major.ps} (symbols).  The solid lines
show the variation of optical depth in a spherical cloud model with a
power law density distribution, $\rho\propto r^{-\alpha}$.}
\label{HH108MMS_tau_major.ps}
\end{figure}

\begin{figure}[htb]
\centerline{\psfig{figure=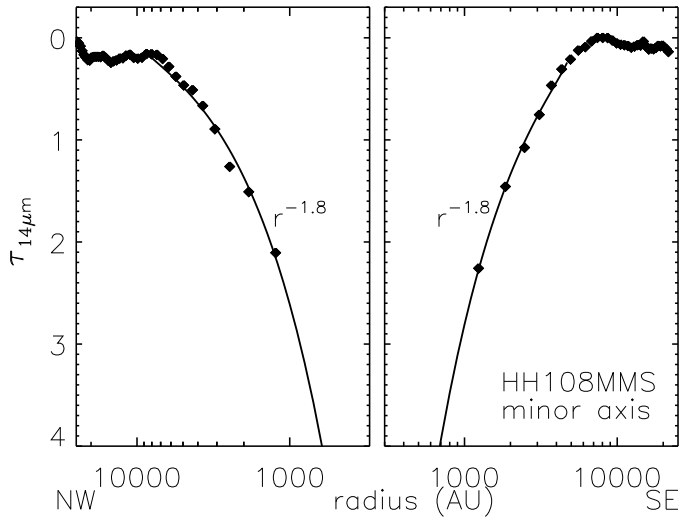,width=10cm,angle=90} }
\caption{ The optical depth profile along the minor 
axis of HH108MMS (symbols).  The solid lines
show the variation of optical depth in a spherical cloud model with a
power law density distribution, $\rho\propto r^{-1.8}$.}
\label{HH108MMS_tau_minor.ps}
\end{figure}

\noindent 
The variation of the optical depth and the best fit density
distributions are shown for the major and minor axis of HH108MMS in
Fig.~\ref{HH108MMS_tau_major.ps} and Fig.~\ref{HH108MMS_tau_minor.ps},
respectively.  In both cuts we derive good fits for a density
distribution $\rho(r) \propto r^{-1.8\pm 0.1}$. The source Q1 is more
difficult to analyze because of the vicinity of HH108MMS and the IRAS
source.

\vskip 1.0cm
\noindent 
{\it 3.3 The spectral energy distribution of IRAS18331--0035} \\

\begin{figure} 
\centerline{\psfig{figure=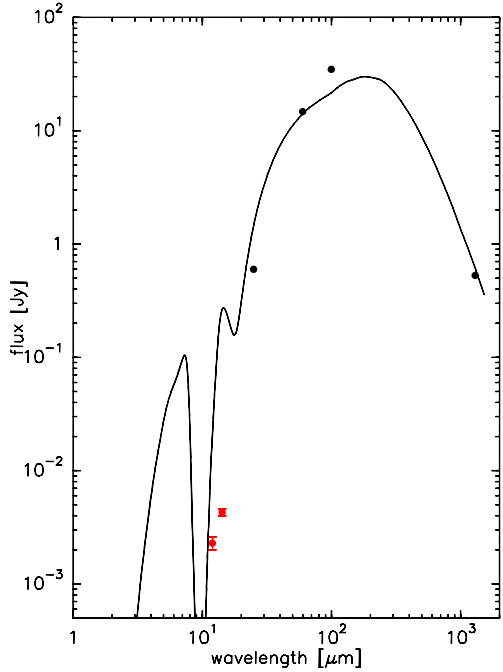,width=7cm} }
\caption{The spectral energy distribution of IRAS18331--0035}
\label{hh108_iras_sed.ps}
\end{figure}

\noindent 
We do a simple spherical radiative transfer calculation for
IRAS18331--0035 to derive estimates for the source structure; the dust
model is from Kr\"ugel \& Siebenmorgen (1994) for fluffy and compact
particles, the numerical code is described in Siebenmorgen et
al.~(1992).  The few photometric points between 12$\mu$m and 1.3mm are
fit by a source with a $1/r$--density distribution.  Its inner and
outer radius are 10$^{12}$ and 10$^{17}$\,cm, there is a visual
extinction to the center of 220\,mag, the total mass and total
luminosity are 1.1\,M$_\odot$ and 2.5\,L$_\odot$, respectively. The
spectral energy distribution is shown in
Fig.~\ref{hh108_iras_sed.ps}. \\

\vskip 1.0cm
\noindent 
{\it  3.4 Mid--infrared polarisation images} \\

\noindent 
Polarisation images are presented in Fig.~\ref{IRAS18331_pol_lw3.ps}
and Fig.~\ref{IRAS18331_pol_lw10.ps}.  They have the following
characteristics:
\begin{itemize}
\item[--] Mid--infrared polarisation is detected, but only along the absorbing
filament.
\item[--] The polarisation is strongly associated with the sources HH108MMS,
Q1 and IRAS18331--0035.
\item[--] The degree of polarisation is high and reaches 50\%.
\item[--] Except for the western part of IRAS18331--0035, there is definitely
some alignment between the polarisation vectors and the filament.
\item[--] Because of their low temperatures, HH108MMS and Q1 do not emit at
14$\mu$m, so polarised emission can be ruled out.
\item[--] Because the background light source, as viewed from us, is behind the
protostars, we exclude the possibility of polarisation from scattering.
Furthermore, scattering would require micron--sized dust particles. \\
\end{itemize}

\noindent 
{\it 3.5 The degree of polarisation in the mid--infrared} \\

\noindent 
We now estimate whether an extinction optical depth at 14$\mu$m of
order two can polarise the light by 30\% or more.  This has to be
checked because polarisation through extinction in the visible is
always much smaller.  In the visible, we have the empirical relation
$p_{\rm V} \le 0.03\,\tau_{\rm V}$ between the visual optical depth
$\tau_{\rm V}$ and the degree of polarisation $p_{\rm V}$.

\noindent 
We therefore compute the cross section of spheroids in the Rayleigh limit when
the wavelength is much bigger than the particle.  A spheroid has three major
axes, $a\ge b\ge c$.  If it has the shape of a cigar (prolate spheroid), $a\ge
b= c$; if it is like a pancake (oblate spheroid), $a= b\ge c$.  The absorption
cross section at wavelength $\lambda$ of a spheroid of volume $V$ when the
electric field vector is parallel to axis $a$ is
\begin{equation} \label{Gl8}
C_a \ = \ {2\pi V\over \lambda} \cdot
\mbox{Im} \left\{{\varepsilon - 1 \over 1+L_a(\varepsilon -1)} \right\}
\end{equation}

\noindent 
Im$\{z\}$ signifies the imaginary part of a complex number $z$; \ $\varepsilon
(\lambda)$ denotes the complex dielectric permeability; $L_a$ is a shape factor
which lies between zero and one and depends on the eccentricity of the grain
(see, for example, Bohren \& Huffman 1983).  There are
analogous definitions for $C_b$ and $C_c$.  For cigars, $L_a + 2L_c= 1$, for
pancakes, $2L_a + L_c= 1$.

\noindent 
Suppose the spheroids are aligned and rotate about the axis of
greatest moment of inertia, which is axis $b$ or $c$ for cigars, and
axis $c$ for pancakes.  Let the light propagate in a direction
perpendicular to the rotation axis and let $C_{{\bf
E}\;\parallel\;{\rm rot}}, C_{{\bf E}\,\perp\,{\rm rot}}$ denote the
time--averaged cross sections in the case when the electric vector of
a linearly polarised incident wave is parallel and perpendicular to
the rotation axis, respectively.

\noindent 
For spinning cigars, the cross section changes periodically.  Because
the mean of $\cos^2 x$ over a rotation cycle is one half, one finds
\begin{equation}\label{Gl9}
C_{{\bf E}\,\perp\,{\rm rot}} = \mbox{${1\over 2}$} \, [C_a + C_c] \ ,
\qquad C_{{\bf E}\;\parallel\;{\rm rot}} = C_b \ .
\end{equation}

\noindent
If a pancake rotates about its axis of maximum moment of inertia,
which is $c$, no averaging is necessary and
\begin{equation}\label{Gl10} 
C_{{\bf E}\,\perp\,{\rm rot}} = C_a \ , \qquad C_{{\bf E}\;\parallel\;{\rm
rot}} = C_c \ .
\end{equation}

\noindent 
The optical depth $\tau$ of a cloud for linearly polarised light
changes with the direction of the electric field vector.  Let
$\tau_{\rm max}$ be the maximum value of $\tau$ and $\tau_{\rm min}$
its minimum.  The corresponding electric field vectors are
perpendicular to each other.  For both cigars and pancakes, $\tau_{\rm
max}$ is proportional to $C_{{\bf E}\,\perp\,{\rm rot}}$ and
$\tau_{\rm min}$ to $C_{{\bf E}\;\parallel\;{\rm rot}}$ with the same
proportionality factor.

\noindent 
When unpolarised background light of intensity $I_*$ traverses the cloud, it is
weakened to the observed intensity
\begin{equation}\label{Gl11}
I_{\rm obs}\ =\ I_*\cdot {e^{-\tau_{\rm max}}+e^{-\tau_{\rm min}}\over 2} \ .
\end{equation}

\noindent
Putting $I_{\rm obs} = I_* \, e^{-\tau_{\rm eff}}$ defines an effective optical
depth
\begin{equation}\label{Gl12}
\tau_{\rm eff} \ = \ -\ln {e^{-\tau_{\rm max}} + e^{-\tau_{\rm min}} \over 2}
\ > \ 0 \ .
\end{equation}

\noindent
The degree of polarisation is obviously defined by
\begin{equation}\label{Gl13}
p(\lambda) \ = \ {e^{-\tau_{\rm min}} - e^{-\tau_{\rm max}}\over e^{-\tau_{\rm
max}} + e^{-\tau_{\rm min}}} \ .
\end{equation}

\noindent
In case of weak extinction, $\tau_{\rm eff} = \mbox{$1\over 2$} \,
(\tau_{\rm max} + \tau_{\rm min})$ and $p = \mbox{$1\over 2$}\,
(\tau_{\rm max} - \tau_{\rm min})$.  If $\tau_{\rm max}-\tau_{\rm
min}$ is large, the polarisation goes to unity.

\begin{figure} 
\vspace{-3.5cm}\centerline{ \psfig{figure=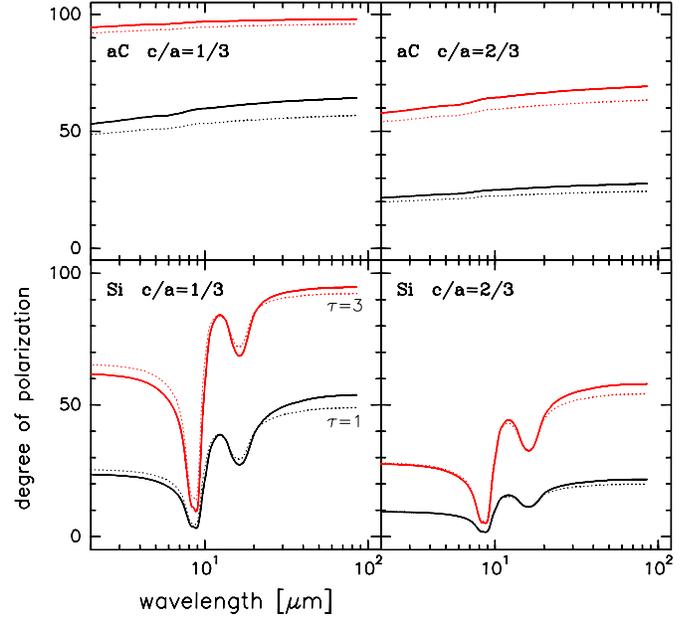,width=9cm}}
\caption{The percentage of polarisation due to perfectly aligned
spinning spheroids of silicate and amorphous carbon of axial ratio
$c/a=1/3$ and $c/a=2/3$.  The effective optical depth $\tau_{\rm
eff}$ has the same value {\it at all wavelengths}.  Solid lines refer
to cigars, dotted to pancakes; in each box, the lower pair of curves
to $\tau=1$, the upper to $\tau=3$.  The size of the grain is
irrelevant as long as it is much smaller than the wavelength.  }
\label{f16k8.ps}
\end{figure}

\noindent
In Fig.~\ref{f16k8.ps} we calculate $p(\lambda)$ assuming an
effective optical depth $\tau_{\rm eff}$ at all wavelengths of either
1 or 3.  The frequency dependent dielectric permeability for silicate
is from Laor \& Draine (1993) and for carbon from Zubko et al. (1996,
their type BE).  We see that with an effective optical depth of two,
even particles of very moderate elongation ($a/c=3/2$) produce very
substantial polarisation, provided they are well aligned.  Therefore,
the observed high degree of polarisation at 14$\mu$m does not seem to
pose a principal problem.

\section{Discussion: Pre--stellar evolution }

\noindent 
Among many other models, the self--similarity solutions of Shu et al.~(1977) for
the early evolution of protostars predict in the absence of rotation and
magnetic fields a $\rho \propto r^{-2}$ profile in the outer envelope and a less
steep distribution $\rho \propto r^{-3/2}$ in the inner part.  Basu \&
Mouschovias (1995), who include magnetic fields and rotation, also find a power
law density distribution in the envelope with an exponent between $-1.5$ and
$-1.8$ and a constant density central region of $\sim$50\,AU size.  Masunaga \&
Inutsuka (1999) perform calculations for the early collapse (before dissociation
of H$_2$ in the core) of a 1\,M$_\odot$ star.  The luminosity stays below
0.1\,L$_\odot$, there is a core of 5\,AU and the density goes like $r^{-2}$ in
the envelope.  Their models should be applicable to HH108MMS and Q1 of the
present paper which also have no mid--IR embedded source.

\noindent 
In a few cases, the predicted theoretical density profiles could be checked by
1.3mm observations of the dust emission (Ward--Thompson et al.~1994, Andr\'e et
al.~1996, Ward--Thompson et al.~1999) and were generally corroborated.  In this
method, the conversion of fluxes into absolute column densities depends on the
dust absorption coefficient at 1.3\,mm and the grain temperature, whereas the
power law exponent of the density distribution is sensitive to the temperature
gradient.

\noindent 
On the other hand, our derivation of the optical depth and the ensuing
density profile from absorption measurements is temperature
independent.  ISOCAM observations similar to ours were also carried
out by Bacmann et al.~(2000).  For the sources which they present, the
maximum optical depth at the cloud center is much smaller than those
for HH108MMS and Q1, so we probe deeper into the cloud, still the
density profiles in the {\it envelopes} roughly agree with what we
derive.  In the {\it cloud center} ($\le 2000$AU), they find a
flattening of the column density which we do not see.  Such a
flattening is expected to occur in the central 1000AU of an isothermal
sphere with temperature $T \sim 10$K, radius of 10000AU and mass of
1\Msun\ simply by solving the hydrostatic equation (Bonnor
1956). \noindent

\noindent 
The measured mid infrared to millimeter dust extinction ratio of $
\kappa_{14\mu \rm{m}} / \kappa_{1300\mu \rm{m}} \sim 470 $ should be
compared with dust models of protostellar environments.  The ratio is
a factor 4 lower than expected for dust in the diffuse medium and it
indicates that the grains in the dense and cold environment of the
protostar HH108MMS are, as expected (Ossenkopf 1993), of rather fluffy
and composite nature.  The measured mid infrared to millimeter dust
extinction ratio of HH108MMS is already tending towards somewhat
elongated grain structures: In the fluffy composite but spherical dust
model by Kr\"ugel \& Siebenmorgen (1994) a ratio of
$\kappa_{14\mu\rm{m}} / \kappa_{1300\mu \rm{m}} \sim 1000$ is found.
Because elongated particles are much better antennas at 1.3mm, they
can give a lower value.  A proof that the dust in the absorbing cores
and the IRAS source is indeed of elongated structure is found by the
ISOCAM polarisation measurements.

\noindent 
As the filament is seen in absorption against the background and assuming that
the majority of the grains are still sub-micron sized particles, one may neglect
dust scattered light at 14$\mu$m.  

Consequently the most plausible mechanism to produce the polarisation is
dichroism.  For polarisation due to extinction of elongated spinning and aligned
dust particles, the polarisation vectors (Figs.~\ref{IRAS18331_pol_lw3.ps} and
\ref{IRAS18331_pol_lw10.ps}) indicate the magnetic field direction.  It appears
that the magnetic field vectors are roughly aligned with the absorbing
filament.  Such ordered fields are also detected from 850$\mu {\rm m}$ maps of
thermal dust emission of prestellar cores (Ward-Thompson et al.~2000) and FIR
polarimetry on the protostar IRAS20503+6006 (Clements et al.~1999).


\section {Conclusion}

\noindent 

\noindent 
We study the protostellar system HH108MMS in the mid IR with
ISOCAM. The 14.3$\mu$m (lw3) image reveals an extended (200$'' \times
20''$) structure which is seen in absorption against the diffuse
background radiation.  Within it, we detect IRAS18331--0035 emitting a
flux of a few mJy and at least two absorbing cores, of which one
coincides with the protostar HH108MMS.

\noindent 
We transform the 14$\mu$m surface brightness of the absorbing cores into
optical depth profiles from which we derive the density structure of the protostars. Along the minor and major axis of
the absorbing cores the density profile is consistent with a $r^{-1.8}$ power
law. 

\noindent 
From the comparison to the 1.3\,mm map (Chini et al.~1997) we derive
observationally an extinction ratio $\kappa_{14\mu\rm{m}} / \kappa_{1300\mu
\rm{m}} \sim 470$.  When compared to dust models (Kr\"ugel \& Siebenmorgen
1994), this ratio points towards rather fluffy and elongated grain structures,
as expected for cold and dense environments.

\noindent 
The structure of IRAS18331--0035 is estimated by applying a radiative
transfer model to fit its spectral energy distribution.  A reasonable
fit is obtained using a $1/r$ density structure, a total cloud mass of
$M_{\rm gas} = 1.$\,1\Msun\ and a luminosity of $L = 2.5$\,\Lsun. The
model envelope has a visual extinction of $A_{\rm V} \sim 220$mag,
where we measure $A_{\rm V}$ from the cloud surface to the inner loci
where grains are evaporating.

\noindent 
We also present the first mid infrared polarisation images of protostellar
absorbing cores prior to the formation on a luminous shock front.

\noindent 
The polarisation pattern is well correlated with the triplet system
and follows roughly the absorbing filament.  The most plausible
explanation for the polarisation is extinction of the background
radiation by elongated spinning and aligned dust particles.  Therefore
the polarisation vectors indicate the magnetic field structure of the
protostellar triplet system.

\noindent 
We detect a fractional polarisation which is much higher ($\geq$15\%) than known
for optical wavelengths ($<10\%$). For various kinds of rotating dust particles
it is shown that dichroic polarisation can indeed produce such high degree of
mid infrared polarisation.

\acknowledgements 
We are grateful to L.~Metcalfe for supporting us with his temporal flat fielding
method. CIA is a joint development by the ESA Astrophysics Division and the
ISOCAM Consortium. The ISOCAM Consortium is led by the ISOCAM PI, C.~Cesarsky.

\end{document}